\documentclass[sigconf]{acmart}
\usepackage[utf8]{inputenc}

\PassOptionsToPackage{table,xcdraw}{xcolor} 

\usepackage{textgreek}
\usepackage{multirow}
\usepackage{subcaption}
\usepackage{xcolor,colortbl}
\definecolor{gray}{rgb}{0.1,0.1,0.1}
\usepackage[T1]{fontenc}
\usepackage{graphicx}
\usepackage{tabularx}
\usepackage{longtable}
\usepackage{enumitem}
\usepackage{amsmath}
\usepackage{graphicx}
\usepackage{array}
\usepackage{booktabs}
\usepackage{makecell}
\usepackage{framed} 
\usepackage{ragged2e}
\hypersetup{
  colorlinks=true,
  linkcolor=teal,
  citecolor=teal,
  urlcolor=teal
}

\AtBeginDocument{%
  \providecommand\BibTeX{{%
    \normalfont B\kern-0.5em{\scshape i\kern-0.25em b}\kern-0.8em\TeX}}}

\copyrightyear{2026}
\acmYear{2026}
\setcopyright{cc}
\setcctype{by}
\acmConference[IDC '26]{Proceedings of the 25th Interaction Design and Children Conference}{June 22--25, 2026}{Brighton, United Kingdom}
\acmBooktitle{Proceedings of the 25th Interaction Design and Children Conference (IDC '26), June 22--25, 2026, Brighton, United Kingdom}
\acmDOI{10.1145/3773077.3816192}
\acmISBN{979-8-4007-2283-7/2026/06}

\begin{document}
\title{Sustainable Care: Designing Technologies That Support Children's Long-Term Engagement with Social Issues}

\author{JaeWon Kim}
\affiliation{%
  \institution{University of Washington}
  \city{Seattle}
  \country{USA}}
\orcid{0000-0003-4302-3221}
\email{jaewonk@uw.edu}

\author{Aayushi Dangol}
\authornote{Both authors contributed equally to this research.}
\affiliation{%
  \institution{University of Washington}
  \city{Seattle}
  \state{WA}
  \country{USA}}
\email{adango@uw.edu}

\author{Rotem Landesman}
\authornotemark[1]
\affiliation{%
  \institution{University of Washington}
  \city{Seattle}
  \state{WA}
  \country{USA}
}
\email{roteml@uw.edu}

\author{Alexis Hiniker}
\affiliation{%
  \institution{University of Washington}
  \city{Seattle}
  \country{USA}}
\email{alexisr@uw.edu}

\author{McKenna F. Parnes}
\affiliation{%
  \institution{Seattle Children's Research Institute}
  \city{Seattle}
  \country{USA}}
  \orcid{0000-0003-2974-9712}
\email{mckenna.parnes@seattlechildrens.org}

\renewcommand{\shortauthors}{JaeWon Kim, et al.}

\begin{abstract}
Children today encounter social issues---climate change, conflict, inequality---through digital technologies, and the design of that encounter shapes whether young people move toward lasting civic engagement or toward anxiety and withdrawal. Much of the content children see is optimized for attention through fear and urgency, with few pathways toward meaningful action---contributing to rising distress and disengagement among young people who care deeply but feel powerless to act. This half-day workshop introduces ``sustainable care'' as a design lens, asking how technology might support children's sustained engagement with social causes without contributing to empathic distress or burnout. We invite researchers and practitioners across child-computer interaction, games, education, and youth mental health to map this landscape together and develop a research agenda for the CCI community.
\end{abstract}



\maketitle

\section{Background}
Technology shapes how young people encounter social issues---and whether that encounter leads to lasting engagement or to overwhelm. Children today are more aware of global challenges than any previous generation~\cite{rideout2019teens, surgeongeneral2021}, exposed daily to climate change, political conflict, social inequality, and humanitarian crises through social media, news apps, AI companions, and online communities. This awareness can spark political interest, strengthen civic identity, and motivate collective action~\cite{kahne2013youth, boulianne2015social}---but the platforms mediating these encounters prioritize attention capture over constructive engagement, amplifying alarming content~\cite{parnes2024socialmediareview, landesman2024instagram} with documented consequences for youth mental health~\cite{surgeongeneral2021}: rising anxiety and depression~\cite{lewandowski2024climateemotions, davis2025instagramemotions}, existential distress rooted in feeling betrayed by protective institutions~\cite{marks2021climate, hickman2021climate}, and a growing sense of powerlessness~\cite{clayton2021mental, ojala2012hope}. Marks et al.'s ten-country survey found 59\% of young people very or extremely worried about climate change, with 45\% reporting that this distress affects their daily functioning~\cite{marks2021climate}. This generation's awareness, in other words, produces distress rather than agency---a collapse of the conditions for sustained engagement.

This workshop introduces ``sustainable care'' as a lens for understanding and addressing this challenge. Sustainable care refers to ways of engaging with social causes that children can maintain over time without experiencing burnout or emotional exhaustion. The term carries a dual meaning: it encompasses both children's care toward societal issues and our care toward children as designers, educators, and researchers. These two dimensions are inseparable. Many young people respond to the issues they encounter online through civic engagement---organizing around causes they care about, raising awareness among peers, and participating in collective action both online and offline~\cite{boulianne2020youth, schwartz2022climateactivism}. Supporting their well-being and supporting that engagement are not competing goals but complementary ones.

\begin{table*}[!h]
\centering
\small
\caption{This half-day, hybrid workshop brings together CCI researchers, game designers, mental health practitioners, civic technology developers, and educators to examine how technology can support sustainable care for young people engaging with social issues. (Workshop Website: \href{https://sustainable-care-idc.github.io/}{\protect\textcolor{teal}{https://sustainable-care-idc.github.io/}})}
\begin{tabular}{@{}p{1.4cm}p{3.7cm}p{11cm}@{}}
\toprule
\textbf{Time} & \textbf{Activity} & \textbf{Description} \\
\midrule
09:00--09:10 & Welcome \& Overview & Agenda, goals, and introductions to the day \\
09:10--09:30 & Participant Introductions & Name, context, and what you hope to take away \\
09:30--09:40 & Framing Sustainable Care & Introduction to the workshop theme and four themes \\
09:40--10:00 & Individual Reflection & Recall a moment of ``unsustainable care''---firsthand or observed---then share in small groups \\
\midrule
10:00--10:45 & \textbf{Roundtable 1:} By Topic Area & Grouped by thematic interest to identify patterns of unsustainable care and the role of children \\
10:45--11:00 & Report-Back & Each group shares key takeaways \\
11:00--11:20 & \textit{Break} & \\
\midrule
11:20--12:05 & \textbf{Roundtable 2:} By Technology & Grouped by technology of interest to examine how specific tools mediate unsustainable care \\
12:05--12:15 & Report-Back & Each group shares key takeaways \\
\midrule
12:15--12:55 & \textbf{Roundtable 3:} Action Planning & Return to Roundtable 1 groups; synthesize insights from both rounds and develop concrete action plans through the four themes of sustainable care \\
12:55--13:10 & Report-Back & Each group shares action plans \\
13:10--13:30 & Closing & Feedback, next steps, and follow-up process \\
\bottomrule
\end{tabular}
\label{tab:schedule}
\end{table*}

This workshop proposes four themes for sustainable care, derived from the US Surgeon General's advisory on protecting youth mental health~\cite{surgeongeneral2021}. The advisory's eight recommendations for young people map onto four broader themes that structure our workshop discussion.
\begin{enumerate}
    \item \textbf{\textit{Bounded responsibility}}: understanding one's role as part of a collective effort rather than carrying the weight of global problems alone. This theme draws on the advisory's recommendations to ``be intentional about your use of social media, video games, and other technologies'' and to ``take care of your body and mind.''
    \item \textbf{\textit{Actionable pathways}}: having concrete, age-appropriate ways to contribute that connect awareness to meaningful action. This theme draws on ``find ways to serve'' and ``be a source of support for others.''
    \item \textbf{\textit{Resilience through community}}: developing capacities to persist through setbacks and uncertainty without losing hope or motivation. This theme draws on ``ask for help'' and ``invest in healthy relationships.''
    \item \textbf{\textit{Mental health orientation}}: recognizing mental health both as a precondition for and a desired outcome of sustainable care. This theme draws on ``remember that mental health challenges are real, common, and treatable'' and ``learn and practice techniques to manage stress and other difficult emotions.''
\end{enumerate}
These components are not a rigid framework but rather a starting point for discussion---a vocabulary for asking what sustainable care might look like in practice.

For the child-computer interaction community, this framing opens important design questions. Technologies can exacerbate the awareness-agency mismatch through fear-based content, algorithmic amplification of distress, and passive consumption patterns~\cite{kim2025privacynorm, radesky2022moralpanic, lukoff2018mindless, dangol2025genai, kim2026socialmediafeellike}. But technologies can also support sustainable care---through games that model civic engagement, tools that connect children to communities of practice, or platforms that make pathways visible and achievable~\cite{kim2024positech, parnes2025digitalinnovations, landesman2023letkids, dangol2025reading, kim2025designforhope}. The question is not whether children should encounter difficult realities but how technology might mediate that encounter in ways that sustain rather than deplete their capacity to care.

Table~\ref{tab:schedule} outlines the half-day workshop structure. A pre-workshop survey collects each participant's technology and topic interests to pre-assign groups for the design activities. The session moves from individual reflection to shared vocabulary building, then through three structured roundtables---organized first by thematic interest, then by technology, and finally returning to thematic groups to map insights onto the four themes of sustainable care and develop concrete action plans (e.g., research collaborations, strategies for embedding sustainable care in ongoing work, draft guidelines). Following the workshop, we will share outputs that participants collectively agree to make public and follow up to support the plans developed during the session.


\section{Organizers}
\textbf{JaeWon Kim} is a PhD candidate at the University of Washington Information School. Her research focuses on understanding, designing, and building social technologies that center on meaningful social connections, especially for youth.

\textbf{Aayushi Dangol} is a PhD candidate at the University of Washington. Her research investigates how to responsibly deploy AI in ways that support children's learning, development, and interaction.

\textbf{Rotem Landesman} is a PhD candidate at the University of Washington Information School. Her research focuses on supporting youth's critical and ethical thinking about emerging technologies.

\textbf{Alexis Hiniker} is an Associate Professor at the University of Washington Information School. She studies how attention-economy design exploits users of all ages---but particularly children, teens, and families---and she designs more respectful alternatives to help people thrive.

\textbf{McKenna Parnes} is an Assistant Professor in the Department of Pediatrics at the University of Washington School of Medicine and an Investigator in the Treuman Katz Center for Pediatric Bioethics and Palliative Care at Seattle Children's Research Institute. She studies how socio-ecological factors play critical roles in youth risk, resilience, and opportunities, and she evaluates the implementation and effectiveness of resilience-building interventions.




\section{Call for Participation}

We invite researchers, designers, and practitioners to join a half-day workshop on \textit{sustainable care}---supporting children's lasting engagement with social causes through technology design. We welcome position papers (2--4 pages, ACM SIGCHI format) addressing empirical findings on how children encounter social issues through technology; design cases of technologies that support or undermine sustained engagement; perspectives from games, education, mental health, or civic technology; and encore submissions of previously published work with a short statement on relevance to sustainable care. Submissions will be reviewed for relevance and diversity of disciplinary background. Submissions should not be anonymized, and at least one author of each accepted paper must attend the workshop, and all participants must register for the workshop.

\begin{acks}
JaeWon Kim would like to acknowledge the CERES Network, University of Washington Global Innovation Funds (GIF), and Student Technology Funds (STF), which provided support for this work. This work was also funded in part by the Paul G. Allen School of Computer Science \& Engineering Endowed Fund for Excellence and a gift from Google. Alexis Hiniker is a special government employee for the Federal Trade Commission. The content expressed in this manuscript does not reflect the views of the Commission or any of the Commissioners.
\end{acks}

\bibliographystyle{ACM-Reference-Format}
\bibliography{references1, references2, references3, newrefs}

\end{document}